# Experimental observation of non-Abelian topological charges and bulk-edge correspondence

Qinghua Guo[1†], Tianshu Jiang[1†], Ruo-Yang Zhang[1†], Lei Zhang[1,2,3], Zhao-Qing Zhang[1], Biao Yang[1,4*], Shuang Zhang[5*], C. T. Chan[1*]

[1]*Department of Physics and Institute for Advanced Study, The Hong Kong University of Science and Technology, Hong Kong, China*
[2]*State Key Laboratory of Quantum Optics and Quantum Optics Devices, Institute of Laser Spectroscopy, Shanxi University, Taiyuan 030006, China*
[3]*Collaborative Innovation Center of Extreme Optics, Shanxi University, Taiyuan 030006, China*
[4]*College of Advanced Interdisciplinary Studies, National University of Defense Technology, Changsha 410073, China*
[5]*School of Physics and Astronomy, University of Birmingham, Birmingham B15 2TT, United Kingdom*

[†]These authors contributed equally to this work.
*Correspondence to: yangbiaocam@nudt.edu.cn; s.zhang@bham.ac.uk; phchan@ust.hk;

**Abstract:**

In the past decades, topological concepts have emerged to classify matter states beyond the Ginzburg-Landau symmetry breaking paradigm. The underlying global invariants are usually characterized by integers, such as Chern or winding numbers. Very recently, band topology characterized by non-Abelian topological charges has been proposed, which possess non-commutative and fruitful braiding structures with multiple (>1) bandgaps entangled together. Despite many potential exquisite applications including quantum computations, no experimental observation of non-Abelian topological charges has been reported. Here, we experimentally observe the non-Abelian topological charges in a PT (parity and time-reversal) symmetric system. More importantly, we propose non-Abelian bulk-edge correspondence, where edge states are found to be described by non-Abelian charges. Our work opens the door towards non-Abelian topological phase characterization and manipulation.

Topological band theory describes the global topological structure underlying various physical systems, where the interplay between intrinsic spin and external symmetries collectively determines distinct symmetry protected topological phases[1-7]. Different topological phases are characterized by bulk topological invariants, such as Chern and winding numbers[8-11]. Until now, most of these are classified into $\mathbb{Z}$ or $\mathbb{Z}_2$ classes, which are the Abelian groups. The topological charges are commutative and exhibit additive properties. The induced bulk-edge correspondence[1,12,13] also inherits the Abelian nature, with the number of edge states being the difference between the two topological invariants across the domain boundary. This simple yet elegant topological relation serves as an important signature for the prediction and characterization of various topological phases[14-16].

Very recently, it is found that symmetry protected topological phases can go beyond the Abelian classifications[17-19]. With multiple bandgaps entangled together, the underlying topological invariants are represented by non-Abelian groups, which reveal the underlying braiding topological structures. This leads to interesting observables such as trajectory-dependent Dirac node collisions in two-dimensional systems, and admissible nodal line configurations in three-dimension [17,19,20]. Moreover, non-Abelian systems are believed to exhibit many potential applications, such as implementation of fault-tolerant quantum computations[21].

However, the experimental observation of non-Abelian topological charges describing band braiding still remains elusive. And bulk-edge correspondence as the guide principle in topological matters has not yet been discussed in the non-Abelian topological systems. Here, we experimentally demonstrate a three-band PT (parity and time-reversal) symmetric system and characterize the underlying non-Abelian topological charges in the momentum space via mapping the eigenstate-frame onto a two-sphere ($S^2$). Furthermore, we propose the non-Abelian bulk-edge correspondence, where the edge states manifest the non-Abelian topological features accordingly.

Usually, the combination of P and T operators can be represented by complex conjugation $K$ when a suitable basis is chosen[22]. Under PT symmetry, the Hamiltonian is hence real at all momenta $k$, i.e., $H(k) = H^*(k)$. Without loss of generality, a three-band Hamiltonian with two complete bandgaps can take the form of $H(k) = \sum_{n=1}^{3} n|u_k^n\rangle\langle u_k^n|$ with $|u_k^n\rangle$ being a real three-dimensional vector. The corresponding state space of the Hamiltonian is then $M_3 = \frac{O(3)}{O(1)^3}$, where $O(N)$ is the orthogonal group. The space of $O(3)$ gives all possible orthogonal triads of unit vectors, whereas the quotient by $O(1)$ group is due to the fact that eigenstates $\pm|u_k^n\rangle$ represent the same states. Applying the fundamental homotopy group theory, one obtains $\pi_1(M_3) = \mathbb{Q}$, where $\mathbb{Q} = (+1, \pm i, \pm j, \pm k, -1)$ forms the non-Abelian quaternion group[17,23,24], consisting of three anticommuting imaginary units satisfying $ij = k, jk = i, ki = j$ and $i^2 = j^2 = k^2 = -1$. Such topological classifications have previously been applied to the real space line defects in biaxial nematic liquid crystals[23]. Figure 1a illustrates all the elements of $\mathbb{Q}$ and their group multiplications are represented via coloured arrows. Each axis of $\mathbb{Q}$ represents an Abelian subgroup. For example, along $(-1, -i, +1, +i, -1)$, the representation is isomorphic to the four-fold rotation ($C_4$) group. However, the multiplications between different axes exhibit non-commutative properties, as indicated by twelve arc-shaped arrows.

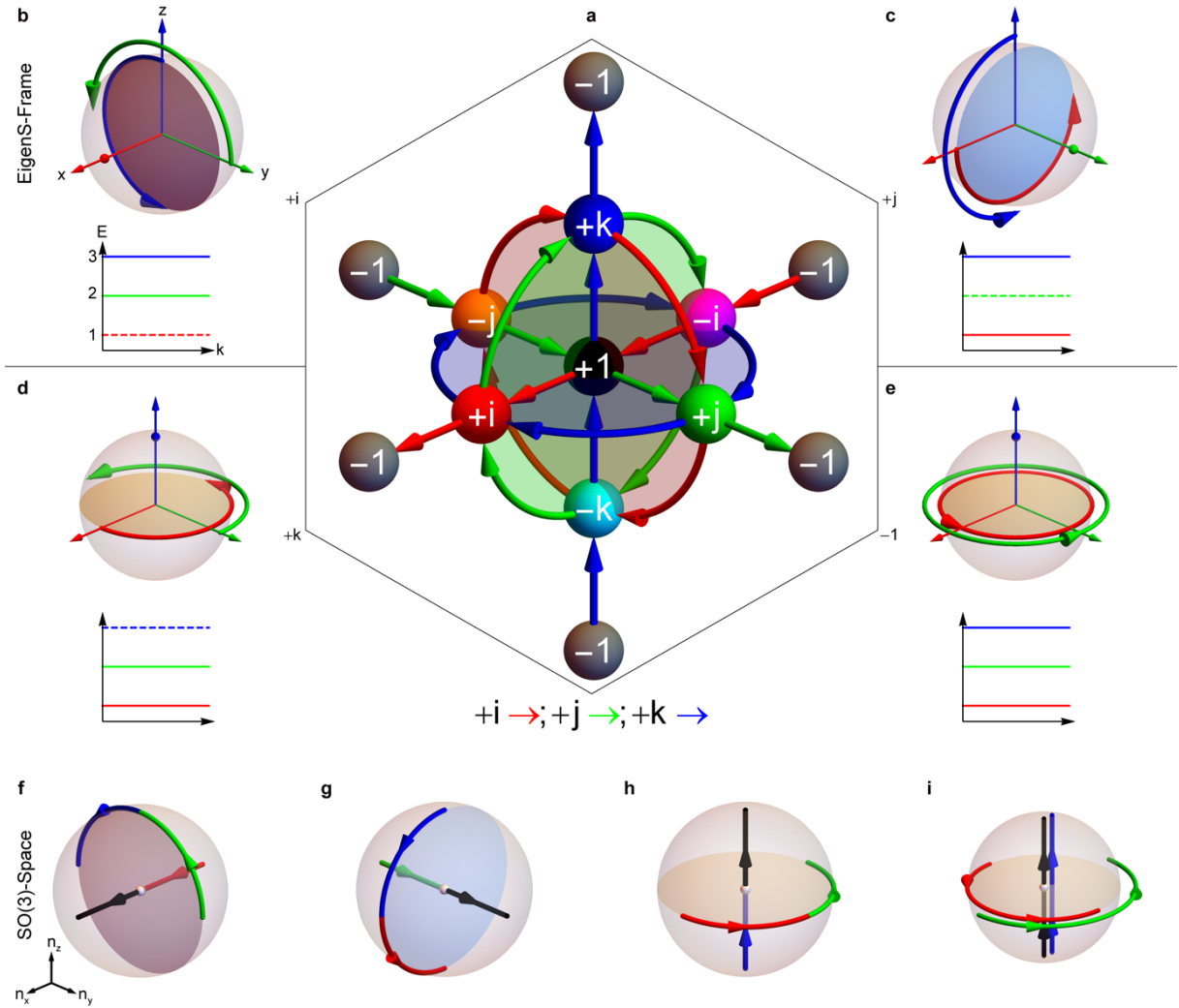

**Figure 1. Non-Abelian topological charges. a.** Graphical representations of the quaternion group $\mathbb{Q}$ and associated non-Abelian structures. The elements are indicated by spheres, with coloured arrows defining their mutual multiplications. Each red/green/blue arrow represents $+i/+j/+k$ multiplication, i.e., $i \cdot (red \rightarrow) = i^2 = -1$, while its reverse direction means multiplying by the corresponding conjugation partner $-i/-j/-k$. **b.** Rotation of eigenstate frame (EigenS-Frame) around the x axis through an angle of $\pi$ induces the non-Abelian topological charge of $+i$. The eigenstate (red dot) of the first band (red dashed line) remains fixed. **c/d.** Similar to (b) with respect to non-Abelian topological charge of $+j/+k$. **e.** Rotation of eigenstate frame around the z axis through $2\pi$ induces the non-Abelian topological charge of $-1$. Red/green/blue colour indicates the real eigenstate trajectory of the $1^{st}$ /$2^{nd}$ /$3^{rd}$ band.

The arrow indicates the direction when $k$ runs from $-\pi$ to $\pi$ across the first Brillouin zone. **f-i.** The curve cosets of $D_2$ in $SO(3)$ space corresponding to (b-e), respectively. Black, red, green and blue curves indicate $R(k)$ acting on $D_2 = [I, R(\hat{x}, \pi), R(\hat{y}, \pi), R(\hat{z}, \pi)]$, respectively. The quaternion group elements can be represented as $1 \to \sigma_0$, $i \to -i\sigma_x$, $j \to -i\sigma_y$ and $k \to -i\sigma_z$, where $\sigma_0$ is the $2 \times 2$ identity matrix and $\sigma_{x,y,z}$ is the Pauli matrix.

We first start with a mathematical model to illustrate the non-Abelian topological charges in a three-band system. The bulk and edge state properties of such non-Abelian systems are subsequently explored theoretically using a tight-binding model and realized experimentally using a braiding network system. For simplicity, we assume the initial eigen-states of the Hamiltonian are three orthogonal vectors as, $H\hat{x} = \hat{x}, H\hat{y} = 2\hat{y}$ and $H\hat{z} = 3\hat{z}$, which form a right-handed Cartesian coordinate frame. At each $k$ we assign a rotation matrix $R(k)$ acting on the initial eigenstate frame, which smoothly rotates the eigenstate frame when $k$ running from $-\pi$ to $\pi$ across the one-dimensional (1D) first Brillouin zone (FBZ). The toy Hamiltonian can then be expressed as, $H(k) = R(k) diag(1,2,3) R(k)^T$ with $R(k) = e^{\vec{\beta}(k) \cdot \vec{L}}$, where $(L_i)_{jk} = -\epsilon_{ijk}$, $\epsilon_{ijk}$ is the fully antisymmetric tensor, and $\vec{\beta}(k) = \phi \hat{n}$ is the product of rotation axis $\hat{n}(k)$ and rotation angle $\phi(k)$. The 1D systems with quaternion charges of $+i, +j, +k$, are achieved via $R(k) = e^{\frac{k+\pi}{2} L_{x,y,z}}$, respectively. The case of $+i$ is shown in Fig. 1b, where the eigenstate frame rotates $\phi = \pi$ around the $x$ axis. It is worth emphasizing that the $x$ axis here can be rotated to other directions via a smooth gauge transformation without topological phase transitions. The inset of Fig. 1b shows three non-degenerate flat bands with the trivial one (the 1st band) marked by dashed lines. Similarly, the other two non-Abelian topological charges $+j$ and $+k$ are given in Figs. 1c and d, where the eigenstate frames rotate $\phi = \pi$ around the $y$ and $z$ axes, resulting in the 2nd and 3rd bands trivial, respectively. The conjugate partners $-i, -j, -k$ correspond to the inverse rotations. From these insets, one can see that the non-Abelian topological phase transitions require band re-ordering, such as the dashed band jumping up/down when the charge alters from $+j$ to $+k/+i$. The process inevitably encounters gap closing and reopening (see supplementary materials, Sec. III, Fig. S5).

On the other hand, the unit element $+1$ represents the trivial case, where the trajectories of the three-eigenstates are continuously contractable to three single points $(1,0,0)$, $(0,1,0)$ and $(0,0,1)$. This is very different from the nontrivial case of charge $-1$ illustrated by Fig. 1e. Taking $R(k) = e^{(k+\pi)L_{x,y,z}}$, the Hamiltonian carries charge of $-1$, corresponding to rotating the eigenstate frame through an angle $\phi = 2\pi$ around $x, y, z$ axis, respectively. It is worth mentioning that one can smoothly transform $R(k)$ between the three cases without closing the gaps (see supplementary materials, Sec. I, Figs. S1, S2 and S6), and as such they all belong to the same topological phase.

It is worth noting that in general, the winding trajectories of eigenstates are not fixed on great circles. In contrast to which axis the states wind about, the crucial property of $\pm i/\pm j/\pm k$ phases is that the trajectories of two of the three bands must terminate at two antipodal points on the sphere, which prohibits the trajectories from contracting to points. And the difference of $\pm i/\pm j/\pm k$ phases is reflected by which two bands are noncontractible. The charge $-1$ corresponds to the case that the eigenstate frame undergoes a rotation of $2\pi$ about an axis through the origin with an arbitrary orientation. The winding trajectories of three bands cannot be contractible simultaneously.

From above we can see that the non-Abelian topological charge is fully determined by the rotation matrix $R(k)$. When $k$ runs from $-\pi$ to $\pi$, $R(k)$ traces out a curve in the parameter space of rotation matrix. One can further understand it via the isomorphism $\frac{O(3)}{O(1)^3} \cong \frac{SO(3)}{D_2}$, where 'SO' indicates the special orthogonal group and $D_2$ is its dihedral point subgroup. One can form the topological space $X/H$ with identifying points of $X$ which can be related by some element of $H$ ($x \equiv xh$). If $R(k) \in SO(3)$, then the trajectories of $R(k: -\pi \to \pi) \circ D_2$ represent the four curve cosets of $D_2$ in the $SO(3)$ parameter space. Here, we depict $SO(3)$ by a solid sphere embedded in the three-dimensional space $\mathbb{R}^3$ with radius of $\pi$. Each point is parametrised by the rotation with normalized axis $\hat{n}(k)$ and rotation angle $\phi(k)$ as $R[\hat{n}(k), \phi(k)] = \exp[\phi(k)\,\hat{n}(k) \cdot \vec{L}]$. It is easy to find that $R(\hat{n}, \pi) = R(-\hat{n}, \pi)$, thus the antipodal points on the surface of the solid sphere are identical. For different non-Abelian charges, the corresponding curve cosets shown in Figs. 1f-i correspond to Figs. 1b-e, respectively. Each (i.e., Fig. 1f) represents four equivalent 1D Hamiltonians up to $D_2$ rotations. These curves all thread a pair of antipodal points, so they are closed and not contractable, which is topologically guaranteed by $\pi_1(SO(3)) = \mathbb{Z}_2$.

Next, we propose a specific tight binding model to realize the aforementioned non-Abelian topological charges, as shown in Fig. 2a. We consider a quasi-1D system with 3 meta-atoms (A, B, C) per unit cell, with inter-cell couplings indicated in Fig. 2a. The real-space Hamiltonian reads,

$$\mathcal{H} = \sum_n \left( \sum_{X=A,B,C} s_{XX} c^\dagger_{X,n} c_{X,n} + \sum_{\substack{X=A,B,C \\ Y=A,B,C}} v_{XY} c^\dagger_{X,n} c_{Y,n+1} + h.c. \right), \quad (1)$$

where $c^\dagger_{X,n}$ and $c_{X,n}$ denote the creation and annihilation operators on the sub-lattice 'X/Y' and site 'n', respectively. After Fourier transformation, we obtain,

$$H(k) = \begin{bmatrix} s_{AA} + 2v_{AA}\cos k & 2u\sin k & 2w\sin k \\ 2u\sin k & s_{BB} + 2v_{BB}\cos k & 2v\sin k \\ 2w\sin k & 2v\sin k & s_{CC} + 2v_{CC}\cos k \end{bmatrix}, \quad (2)$$

where we have set $v_{AB} = v_{BA} = iu$, $v_{BC} = v_{CB} = iv$ and $v_{CA} = v_{AC} = iw$ to make the Bloch Hamiltonian explicitly real ($u, v, w$ are all real). The tight binding model constructed here is PT symmetric despite that it breaks both time-reversal and inversion symmetries. There are 9 independent parameters when only the nearest neighbour (NN) hoppings are considered (see supplementary materials, Sec. II, for the cases involving the next nearest neighbour (NNN) and other hoppings, Fig. S3). Their values uniquely determine the non-Abelian topological charges. We first choose parameters that can mimic the flat-band cases as shown in Figs. 1b-e. In our experiments, those values will be chosen for the ease of implementation. The case of Fig. 1b with the non-Abelian charge of $+i$ can be realized by setting $s_{AA} = 1$, $s_{BB} = s_{CC} = \frac{5}{2}$, $v_{BB} = -v_{CC} = \frac{1}{4}$ and $v = \frac{1}{4}$ (the other parameters are zero unless otherwise specified). The parameters for other cases (Figs. 1c-e) are shown in Tab. S1 (see supplementary materials, Sec. II).

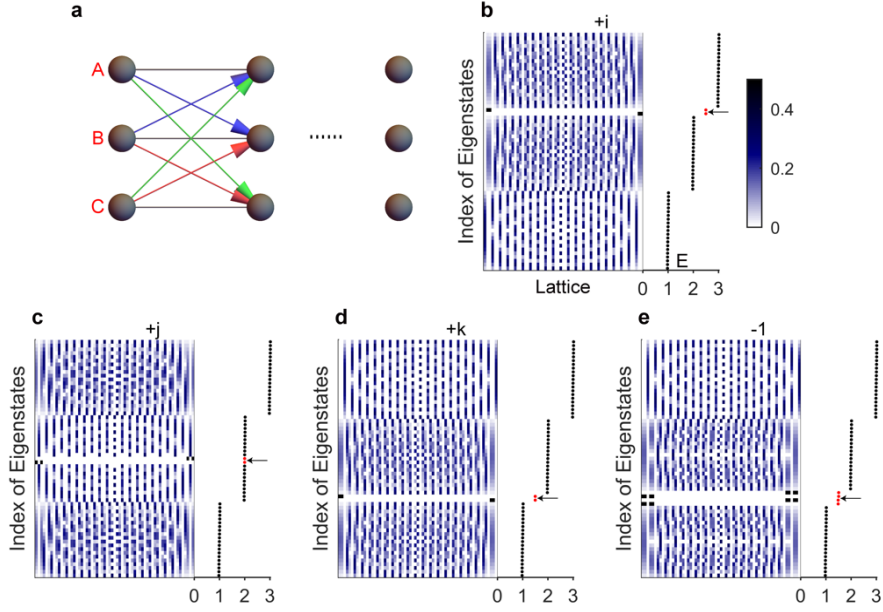

**Figure 2. Realizing non-Abelian topological charges in a tight-binding model. a.** Schematic view of the tight-binding configuration, where only the nearest neighbour (NN) hoppings are considered. The directional hoppings are $v_{AB} = v_{BA} = iu$ (blue), $v_{BC} = v_{CB} = iv$ (red) and $v_{CA} = v_{AC} = iw$ (green). **b-e.** Edge states (red dots) at hard boundaries for systems with charges of $+i$, $+j$, $+k$ and $-1$, respectively. The eigenstate distributions as a function of positions (left panels) are illustrated accompanying with the eigen-energy profiles (right panels). A tiny modulation (1%) is imposed on $v_{XX}$ to avoid flat band. The quasi-1D lattice has 21 periods here.

Figures 2b-e present the edge states localized at the hard boundaries of the quasi 1D lattices with different values of non-Abelian topological charges[17]. The usual Abelian topological charges are used to characterize the topology of a single bandgap, whose value counts the number of edge states inside that gap. However, from the non-Abelian perspective, the non-Abelian charges describe the topology of all gapped bands, and can hence predict both the distribution and number of edge states in all bandgaps. As shown in Fig. 2b, the charge of $+i$ indicates the second gap is nontrivial and supports one edge state on each end of a finite-length

lattice. The 1st band eigenstate is fixed while the 2nd and 3rd band eigenstates rotate through $\pi$ for $k$ running from $-\pi$ to $+\pi$, and the bandgap sandwiched by these two rotating states carries edge states. Similar argument also applies to charge of $+j$ and $+k$ as shown in Figs. 2c and d, respectively. In is interesting to note from Fig. 2c for charge of $+j$ that the edge states merge into the 2nd continuum band, forming bound states in continuum (BICs). It is worth emphasizing that the BICs are not enforced by the non-Abelian topology but accidentally achieved by pertaining to specific parameters. The two edge states with charges of $\pm j$ can be located at any frequency between the 1st and 3rd bands. In contrast, the quaternion charge of $-1$ supports two edge states on each side as shown in Fig. 2e. Nevertheless, the locations of the edge states are fickle and vary according to the details of bulk states[17]. We argue that they have gone beyond the topological description as they all belong to the topological charge of $-1$ but can be continuously transformed between each other (Figs. S6 and S7, see details in supplementary materials, Sec. IV).

In order to observe the non-Abelian topological charges experimentally, we implement the above tight binding model using transmission line networks with braiding connectivity. For a network connected by transmission lines, the wave function of each node satisfies the network equation[25,26]:

$$-\psi_m \sum_n \coth(zl_{mn}) + \sum_n \frac{1}{\sinh(zl_{mn})}\psi_n = 0, \quad (3)$$

where $\psi_m$ and $\psi_n$ are, respectively, wave amplitudes at nodes $m$ and $n$, $l_{mn}$ is the cable length connecting node $m$ and node $n$ with $z = \left(\frac{i\omega}{c_0}\right)\sqrt{\epsilon_r}$ (see details in supplementary materials, Sec. V). This equation form is mathematically equivalent to the zero energy tight binding equation with an on-site term $-\psi_m \sum_n \coth(zl_{mn})$ and hopping term $\sum_n \frac{1}{\sinh(zl_{mn})}\psi_n$. The hopping between two nodes can be realized by connecting these two nodes with transmission lines. Provided that the optical path lengths $L = l_{mn}\sqrt{\epsilon_r}$ in all cables are identical, the network equation can be directly mapped to a tight-binding model. Thus, transmission line network offers an ideal platform to realize various tight binding models.

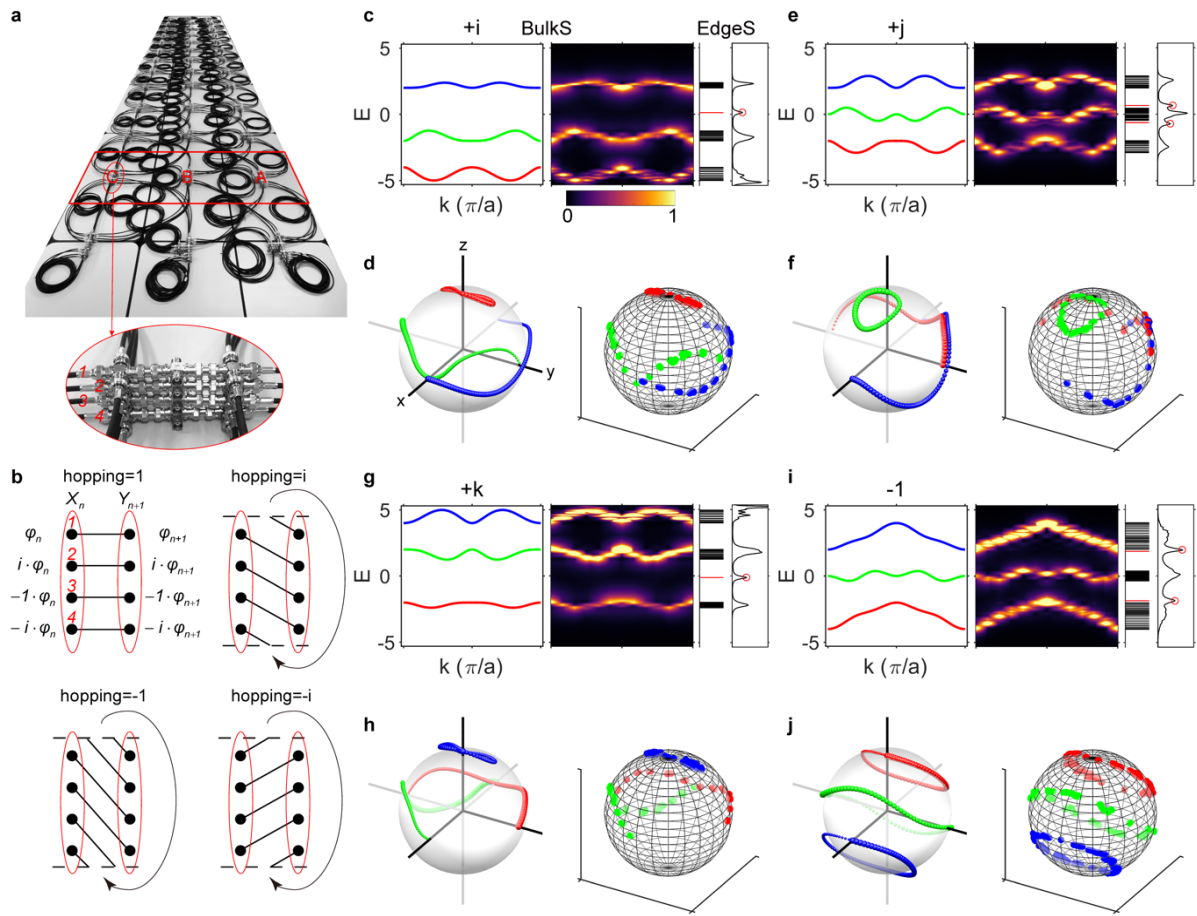

**Figure 3. Experimentally characterize non-Abelian topological charges and hard boundary edge states. a.** The sample photo of transmission line network, where 13 periods are used in the construction. The inset shows each meta-atom consists of four nodes stacked perpendicularly. **b.** Realization of complex number hoppings between meta-atoms $X_n$ and $Y_{n+1}$ with $X$ or $Y = A, B, C$. **c, e, g and i.** Theoretically simulated (left panels) and experimentally mapped (right panels) band structures. The projected edge states are shown in the insets

correspondingly. **d, f, h and j.** Theoretically simulated (left panels) and experimentally mapped (right panels) eigenstate frame spheres. The 1$^{st}$, 2$^{nd}$ and 3$^{rd}$ bands are coloured as red, green and blue, respectively. The direction of line-width decreasing indicates $k: -\frac{\pi}{a} \to \frac{\pi}{a}$ where $a = 1$.

The tight-biding model in the previous section looks deceptively simple but its experimental realization is a bit tricky as it requires complex number hoppings. To realize the complex hoppings, the network is required to have multiple nodes in each meta-atom[27]. Each meta-atom is actually the compactification of a hidden dimension carrying four nodes (layers) from top to down as shown in Fig. 3b. Due to the braiding connectivity, hopping between nodes in neighbouring layers provides a phase shift of $e^{i\pi/2}$. Therefore, four hopping values: $1, i, -1, -i$ are realized here (Fig. 3b).

In the experiment, a network consisting of 13 periods was designed to characterize the non-Abelian topological charges corresponding to $+i, +j, +k$ and $-1$, respectively. Figure 3a is the photo of the sample for the configuration of charge $-1$, where there are three meta-atoms A, B and C in one period and the inset shows that there are four nodes *1, 2, 3, 4* in each meta-atom. The AC frequency signal from a signal generator was injected to excite the eigenstates of the networks. Both the amplitude and phase of the voltage at each meta-atom are collected by an oscilloscope (see detailed methods in supplementary materials, Sec. VI and Fig. S10). After Fourier transformation, we obtain the eigenstate distributions in momentum space. Subsequently the band structure and eigenstates corresponding to each non-Abelian case can hence be mapped experimentally. Figures 3c-j show the experiment results (right panel) which agree well with the tight bonding model results on the left panel. The parameters for each case are summarized in Tab. S2 (see supplementary materials, Sec. II and Fig. S9 for corresponding network connectivity). The insets in Figs. 3c-j show the measured edge state distributions in corresponding band gaps with the theoretical results shown on the left for comparison. Taking the charge $+i$ configuration for example, a peak is observed in the gap between the 2$^{nd}$ and 3$^{rd}$ bands. The theoretical field distributions of the edge states are shown in Fig. S4 for reference (see supplementary materials, Sec. II). Most edge states in Figs. 3c-j can be well understood according to the analysis in Fig. 2. Except that the edge states of charge $+j$ in Fig. 3e do not appear as BICs as shown in Fig. 2c. As we mentioned above, the presence of BICs is special to the flat band situation where the 2$^{nd}$ band states do not couple with the two winding states, while in general cases, the edge states can be anywhere between the 1$^{st}$ and 3$^{rd}$ bands for the charge of $\pm j$. In addition, the edge states for the charge of -1 are fickle as mentioned above.

When two semi-infinite lattices carrying different non-Abelian topological charges form a domain wall, i.e., the charge pair of $(+i, +j)$, some boundary modes confined at the domain wall should emerge. The well-known Abelian bulk-edge correspondence, given by the difference of two integers (Fig. 4a, $\Delta N = N_L - N_R \in \mathbb{Z}$) becomes inadequate for describing the global edge state configuration in non-Abelian topological band systems. Here, we propose a non-Abelian bulk-edge correspondence as schematically shown in Fig. 4a, which claims that the topology of the domain wall is characterized by the relation of $\Delta Q = Q_L/Q_R \ (\in \mathbb{Q})$. From the perspective of group theory, the corresponding domain wall topology of the Abelian ($\Delta N = N_L - N_R$) and non-Abelian ($\Delta Q = Q_L/Q_R$) cases is consistent, because both can be unified as the left charge multiplying the inverse of the right charge. Thus, our non-Abelian bulk-edge correspondence is an extension the Abelian case. As mentioned earlier, the domain wall charge $\Delta Q$ not only determines the number of edge states, but also specifies the arrangement of edge states in each bandgap.

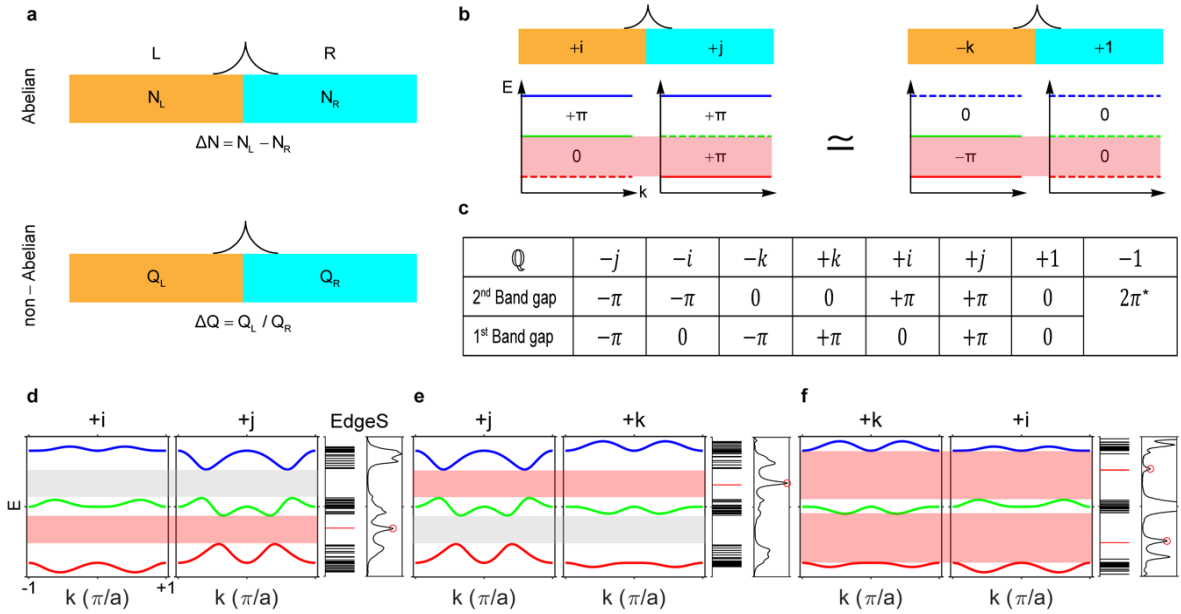

**Figure 4. Non-Abelian bulk-edge correspondence. a.** Schematically illustrating Abelian and non-Abelian bulk-edge correspondence. **b.** Each gap is labelled by the corresponding Zak phase. The first gap carrying different Zak phases across the domain wall supports boundary modes, as shaded in red. The boundary modes are characterized by $\Delta Q = +i/+j = -k/+1$. The boundary states of charge pair $(-k, +1)$ can be regarded as hard boundary edge states of charge $-k$. **c.** Non-Abelian topological charges are related to two-bandgap Zak phases. * The Zak phase in the charge $-1$ is only well defined in special situations when one band is fully decoupled with the other two. **d-f.** Theoretically simulated (left panels) and experimentally measured (right panels) boundary states between two different non-Abelian topological charges.

In order to verify the non-Abelian bulk-edge correspondence, we start from an Abelian perspective. As shown in Fig. 4b, two bandgaps are individually labelled by the corresponding Zak phases. The Zak phase of the 1st bandgap is accumulated via parallel-transporting the 1st band eigenstate, while the Zak phase of the 2$^{nd}$ bandgap is obtained via summing over that of the 1$^{st}$ and 2$^{nd}$ bands. The Zak phase for a single bandgap can only take the value of 0 or $\pi$ as the homotopy mapping $\pi_1\left(\frac{O(3)}{O(2)\times O(1)}\right) = \pi_1(\mathbb{R}P^2) = \mathbb{Z}_2$, which means one cannot distinguish $+\pi$ from $-\pi$. However, because the system is fully described by the non-Abelian topological charges where the rotation of the eigenstate frame having the fixed direction, so we can uniquely label each gap with $-\pi$, 0 and $+\pi$ respectively. In other words, the system possesses a more refined topological structure than $\mathbb{R}P^2$, namely $M_3$ as defined earlier. The full labelling list is shown in Fig. 4c. There are total of 8 different combinations. For the charge $-1$, although we could give the exact Zak phase for each gap in some special cases when one of band is fully decoupled, we cannot determine the Zak phase when it involves arbitrarily as stated in supplementary materials (Sec. IV).

Now, we employ Zak phase to predict the existence of edge states for each single bandgap, although the global picture can only be captured by non-Abelian topological charges. As shown in Fig. 4b for the bandgap shaded in red, the Zak phases change across the domain wall, which implies the presence of topological phase transition and thus the emergence of edge states in the 1$^{st}$ band gap. The boundary mode configuration exactly fits to the hard boundary edge states with charge of $\pm k$ (Fig. 2d). On the other hand, the global analysis based on the non-Abelian bulk-edge correspondence shown in Fig. 4a, predicts that the number and position of domain wall states for the charge pairs of $(+i, +j)$ and $(-k, +1)$ should be the same. This illustrates the non-Abelian bulk-edge correspondence. One can apply similar arguments for all other cases, which are all consistent with $\Delta Q = Q_L/Q_R$. As one can not specify the orientation of the domain wall, that means the charge pairs of $(+i, +j)$ and $(+j, +i)$ take the same edge state distributions. In total, there are five types of edge state distributions, homomorphically corresponding to the five classes $(+1, \pm i, \pm j, \pm k, -1)$ in the quaternion group $\mathbb{Q}$. In the experiment, we construct several different types of domain walls (Tab. S3 and Fig. S11) and mapped their domain wall states accordingly as shown in Figs. 4d-f. The experimental results agree with the theory predictions, further confirming the proposed non-Abelian bulk-edge correspondence. In the domain walls involving charge $-1$, such as the charge pair $(+i, -i)$, the boundary states share similar features with the hard boundary edge states in the configuration with charge $-1$ (see supplementary materials, Sec. VII, Fig. S12).

Our work serves as the experimental observation of the non-Abelian topological charges in the momentum space. The network system provides an easily controllable and reconfigurable platform for studying various topological braiding structures. In addition to demonstrating the non-Abelian bulk topological charges, we propose non-Abelian bulk-edge correspondence by considering the presence of edge states in relation to non-Abelian domain wall charges. The proposed ideas can be implemented in optical regime, where coupled waveguides[28-31] can support non-Abelian photonic states and even realize non-Abelian photon pumping by introducing spatial modulations along the waveguides[32]. Other systems including dynamic optical lattices are also very attractive[33].

**Acknowledgements**

This work is supported by the Hong Kong RGC (AoE/P-02/12, 16304717, 16310420) and the Hong Kong Scholars Program (XJ2019007). L.Z. acknowledges National Natural Science Foundation of China with Grant No. 11704232. S.Z. acknowledges support from the ERC Consolidator Grant (TOPOLOGICAL), the Royal Society and the Wolfson Foundation. C.T.C. thanks Prof. R. Cheng for making space available to construct the network.